\documentclass[hidelinks]{article}
\usepackage{arxiv}
\usepackage{listings}
\usepackage{color}
\usepackage{soul}
\usepackage{graphicx}
\definecolor{light-gray}{gray}{0.95}
\usepackage{subcaption}
\usepackage{enumerate}
\usepackage{rotating}
\usepackage{multirow}
\usepackage{amsmath}
\usepackage{tikz}
\usepackage{pgfplots}
\usepackage{algorithm}
\usepackage{algpseudocode}
\pgfplotsset{compat=1.18}
\usepackage{amssymb}

\definecolor{distNormal}{RGB}{65, 105, 225}   
\definecolor{distPoisson}{RGB}{255, 140, 0}   
\definecolor{distUniform}{RGB}{50, 205, 50}   

\newcommand{\errorDistributionChart}[9]{%
\begin{tikzpicture}
\begin{axis}[
    xlabel={#2},
    ylabel={Error (\%)},
    ylabel near ticks,
    ylabel style={yshift=-5pt},
    yticklabel style={inner xsep=1pt},
    ymin=0,
    ymax=#5,
    xmin=0.5,
    xmax=20.5,
    enlarge x limits=false,
    ytick={#6},
    xtick={1,2,3,4,5,6,7,8,9,10,11,12,13,14,15,16,17,18,19,20},
    xticklabels={#1\_1,#1\_2,#1\_3,#1\_4,#1\_5,#1\_6,#1\_7,#1\_8,#1\_9,#1\_10,#1\_11,#1\_12, #1\_13,#1\_14,#1\_15,#1\_16,#1\_17,#1\_18,#1\_19,#1\_20},
    xticklabel style={rotate=45, anchor=east, font=\footnotesize},
    title={#3},
    title style={font=\bfseries\large},
    grid=major,
    grid style={dashed, gray!30},
    legend style={at={(0.02,0.98)}, anchor=north west, nodes={scale=0.8, transform shape}},
    every axis plot/.append style={
        ybar,
        bar width=0.55,
        draw=black,
        solid,
        thick,
        nodes near coords={\pgfmathprintnumber{\pgfplotspointmeta}\%},
        every node near coord/.style={
            font=\tiny,
            color=black,
            anchor=south,
            yshift=1pt
        }
    }
]

\addplot[fill=distNormal, forget plot] coordinates { #7 };

\addplot[fill=distPoisson, forget plot] coordinates { #8 };

\addplot[fill=distUniform, forget plot] coordinates { #9 };

\addlegendimage{area legend, fill=distNormal, draw=black, solid}
\addlegendentry{Normal}

\addlegendimage{area legend, fill=distPoisson, draw=black, solid}
\addlegendentry{Poisson}

\addlegendimage{area legend, fill=distUniform, draw=black, solid}
\addlegendentry{Uniform}

\end{axis}
\end{tikzpicture}%
}

\definecolor{colActual}{RGB}{220, 20, 60}      

\newcommand{\countComparisonChart}[7]{%
    \begin{tikzpicture}
    \begin{axis}[
        width=1.2\columnwidth,
        height=0.6\columnwidth,
        ybar,
        bar width=10pt,
        ymin=0, ymax=#4,
        ylabel={Count},
        xlabel={Attribute Values},
        xtick={1,2,3,4,5,6,7,8,9,10,11,12,13,14,15,16,17,18,19,20},
        xticklabels={#7},
        xticklabel style={rotate=45, anchor=east, font=\footnotesize},
        title style={font=\bfseries\large},
        legend style={at={(0.98,0.98)}, anchor=north east, nodes={scale=0.8, transform shape}},
        nodes near coords,
        nodes near coords style={
            font=\tiny,
            color=black,
            rotate=75,      
            anchor=west     
        },
        enlarge y limits={upper, value=0.2},
        enlarge x limits=0.15,
        grid=major,
        grid style={dashed, gray!30}
    ]

    \addplot[fill=colActual, draw=black] coordinates {#5};
    \addlegendentry{Expected}

    \addplot[fill=#3, draw=black] coordinates {#6};
    \addlegendentry{Actual}

    \addlegendimage{empty legend}
    \addlegendentry{\textbf{Dist: #2}}

    \end{axis}
    \end{tikzpicture}%
}

\usepackage{listings}
\usepackage{xcolor}

\definecolor{jsonKey}{RGB}{163, 21, 21}      
\definecolor{jsonVal}{RGB}{4, 81, 165}       
\definecolor{jsonNum}{RGB}{9, 134, 88}       
\definecolor{jsonKw}{RGB}{0, 0, 255}         

\lstdefinestyle{jsonbox}{
    basicstyle=\ttfamily\scriptsize,
    columns=fullflexible,
    frame=single,
    numbers=none,
    breaklines=true,
    captionpos=b,
    tabsize=2,
    backgroundcolor=\color{white},
    stringstyle=\color{jsonVal},       
    commentstyle=\color{gray},
    morekeywords={true,false,null},
    keywordstyle=\color{jsonKw}\bfseries,
    moredelim=[s][\color{jsonKey}]{"}{:},
    literate=
     *{0}{{{\color{jsonNum}0}}}{1}
      {1}{{{\color{jsonNum}1}}}{1}
      {2}{{{\color{jsonNum}2}}}{1}
      {3}{{{\color{jsonNum}3}}}{1}
      {4}{{{\color{jsonNum}4}}}{1}
      {5}{{{\color{jsonNum}5}}}{1}
      {6}{{{\color{jsonNum}6}}}{1}
      {7}{{{\color{jsonNum}7}}}{1}
      {8}{{{\color{jsonNum}8}}}{1}
      {9}{{{\color{jsonNum}9}}}{1}
}

\usepackage[maxbibnames=2]{biblatex} 
\AtBeginDocument{
  \providecommand\BibTeX{{%
    \normalfont B\kern-0.5em{\scshape i\kern-0.25em b}\kern-0.8em\TeX}}}

\begin{document}



\title{MuSimA: A Tool with Multi-modal Input for Generating Bespoke ABAC Datasets} 

\author{%
  Saket Jha\\
  Indian Institute of Technology Kharagpur, India\\
  \texttt{pmsaketjha@gmail.com}   \And
  Karthikeya S. M. Yelisetty\\
  Indian Institute of Technology Kharagpur, India\\
  \texttt{yelisettikarthik0@gmail.com}   \And
  Singabattu Sathya \\
Indian Institute of Technology Kharagpur, India\\
  \texttt{sathyasingabattu2004@gmail.com}   \And
  Shamik Sural \\
  Indian Institute of Technology Kharagpur, India\\
  \texttt{shamik@cse.iitkgp.ac.in} 
}
\date{}
\renewcommand{\headeright}{}
\renewcommand{\undertitle}{}

\maketitle
\begin{abstract}
Recent advances in research on Attribute-based Access Control (ABAC) has led to the development of several ingenious methods for representing and enforcing organizational security policies. However, so far little effort has been spent towards building a tool for generating large-scale synthetic datasets that can be used to test the developed ABAC systems. In this paper, we address this shortcoming by building MuSimA - a web-based tool for generating ABAC datasets with user-specified probability distributions of attribute values. It supports multi-modal input, i.e., users can provide specifications either as a structured JSON file or as a combination of a minimal JSON along with hand-drawn distribution sketches. In the latter case, a Large Language Model is used to automatically extract appropriate distribution parameters from the sketches. The generated synthetic ABAC data matching the input specifications can be downloaded by the user. For studying scalability of algorithms and methods related to ABAC, data can be generated for varying sizes and complexities. We make MuSimA freely available for use by the research community. 
\end{abstract}


\section{Introduction}
\label{sec:intro}
Over the last few years, Attribute-based Access Control (ABAC) has emerged as one of the most researched topics in access control. Efforts have been made to improve ABAC policy mining algorithms \cite{ransam}\cite{DBLP:conf/lopstr/BambergerF23}\cite{ABAC_PolicyMining2024}\cite{ABAC_PolicyMining2025}, extraction of machine enforceable ABAC policies from natural language policies \cite{sonune2025lmntoolgeneratingmachine}\cite{MianSACMAT2025}, and enhancing scalability \cite{DBLP:journals/cn/DingWMD25}. ABAC has also found significant applications in the healthcare domain \cite{abac_health}, university systems \cite{abacuniversity}, IoT \cite{abaciot1}\cite{abac_iothealthcloud}, cloud \cite{Weng2025}\cite{abac_iothealthcloud} and blockchain \cite{abacblockchain}, among others. 

A common observation from the above-mentioned and other related literature is that even after claiming new research breakthroughs in ABAC, the experiments are conducted only on synthetic datasets generated by standalone simulation tools \cite{cosequeue}. The often cited reason is a lack of available real data, since organizations are not willing to share their access control datasets. This reasoning is understandable and, to the best of our knowledge, the only two publicly available real-world datasets are Amazon Kaggle \cite{Amazon_dataset} and Amazon UCI \cite{AmazonUciDataset}, which have been used by several researchers for showing the efficacy of their approaches \cite{polisma}\cite{dlbac}. Since these datasets have a fixed number of entries, an effort was made in recent years to build a tool named ConGRASS that uses Conditional Tabular GAN for generating new realistic data samples from the above two real datasets \cite{Rai2023}. However, while it can output variable sized datasets, each data point is strictly limited by the number of attributes present in the respective original datasets. 

Another source of data that has been used in the ABAC literature is due to Xu and Stoller \cite{xustoller}. Although named as case studies, these datasets are essentially simulated data produced by standalone Java programs. The main drawback with these datasets is that they allow only minor pre-determined variations in the simulation process and hence their scope is rather restricted. Besides, it does not allow generation of data sets that can have different kinds of distributions based on the domain of application. ABAC Lab~\cite{ABACLab} is a repository of policy datasets supporting analysis and benchmarking. However, it essentially facilitates ABAC policy mining with predetermined parameters and does not support user specified attribute distributions.
Further, none of the existing tools can handle environmental attributes - a major differentiating factor of ABAC over other access control models like Role-based Access Control (RBAC) and Discretionary Access Control (DAC). 

It is in this context that we propose MuSimA (\underline{Mu}lti-Modal \underline{Sim}ulator for \underline{A}BAC Systems) - a web based application being made freely accessible to researchers. MuSimA takes parameters like number of users, resources, environmental conditions, number of attributes, their values and distribution of the attribute values over their respective entities (user, resource and environmental condition) as input in a JSON file. An ABAC dataset satisfying such requirements is then generated with names of various entities, attributes, values of attributes, and entity-entity attribute value pair assignments, along with a set of ABAC rules. The authorizations consistent with these ABAC components are also produced. All such constituents of the generated ABAC dataset are combined in a Zip file that can be downloaded in the user's system. The generated datasets are useful for testing across multiple layers of an ABAC system: at the Policy Information Point (PIP) for studying attribute storage strategies, at the Policy Decision Point (PDP) for evaluating and mining policies, and at the Policy Enforcement Point (PEP) for security analysis. Besides providing the desired distributions in the input JSON, these can also be hand-drawn and the corresponding images uploaded. The distributions are automatically extracted from these images using a Large Language Model (LLM). 

MuSimA is a Flask-based lightweight web application accessible from anywhere\footnote{\url{https://facweb.iitkgp.ac.in/~shamik/tool.html#musima}}.

\section{Design and Implementation of MuSimA}
\label{sec:des&impl}
In this section, we present the details of MuSimA, including a system overview, input and output formats, and the data generation process.

\subsection{System Overview}
\label{subsec:sysoverview}

MuSimA follows a systematic approach for generating consistent and domain-specific datasets. A user interacting with the tool first provides a specification that defines the structure and distribution characteristics of the desired ABAC dataset relevant to the domain for which an  algorithm or system is being tested. Based on these inputs, MuSimA produces a complete ABAC dataset with output files representing different views of the generated dataset. The user interface of the web application through which the users interact is minimalistic.

\begin{figure}[h]
    \centering
    \includegraphics[width=0.53\linewidth]{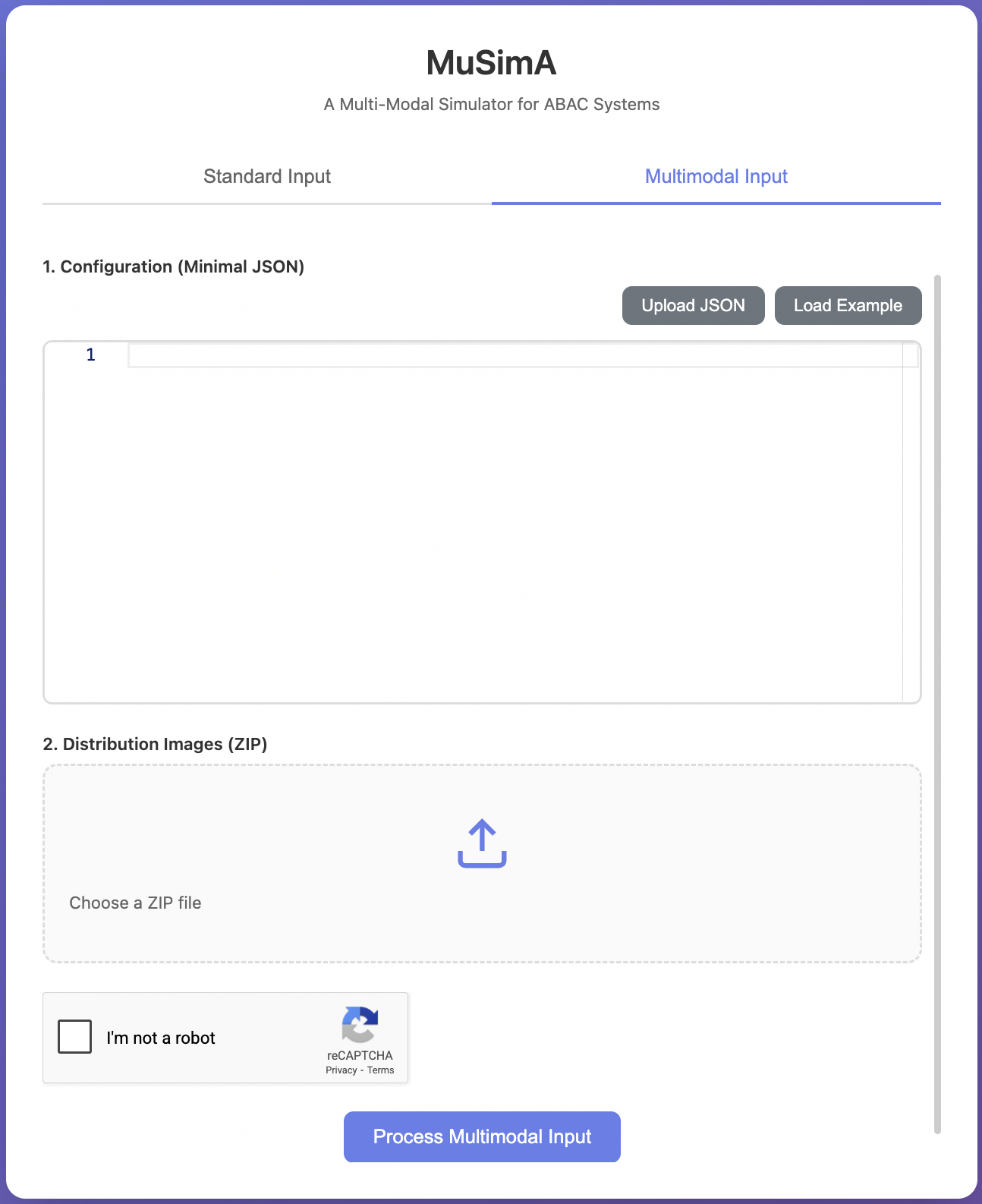}
    \caption{MuSimA Web Interface}
    \label{fig:webinterface}
\end{figure}

The MuSimA tool provides a user-friendly interface for generating synthetic ABAC datasets. The web interface, shown in Figure \ref{fig:webinterface}, allows users to specify configuration parameters either textually or by uploading images.

\begin{lstlisting}[style=jsonbox, caption={Example Input JSON for Data Generation}, label={lst:inputjsonsample}]
{
  "subject_size": 3, "object_size": 3, "environment_size": 2, 
  "permit_rules_count": 1, "deny_rules_count": 1,
  "subject_attributes_count": 2, "object_attributes_count": 2,
  "environment_attributes_count": 1,
  "subject_attributes_values": [2, 4],
  "object_attributes_values": [2, 1],
  "environment_attributes_values": [2],
  "subject_distributions": [
    { "distribution": "U" },
    { "distribution": "N", "mean": 2, "variance": 1 }
  ],
  "object_distributions": [
    { "distribution": "P", "lambda": 1 },
    { "distribution": "U" }
  ],
  "environment_distributions": [{ "distribution": "U" }]
}
\end{lstlisting}


\subsection{Input Specification Format}
\label{subsec:inputspec}

\begin{figure*}[t]
    \centering
    \begin{subfigure}{0.8\textwidth}
    \includegraphics[width=\linewidth]{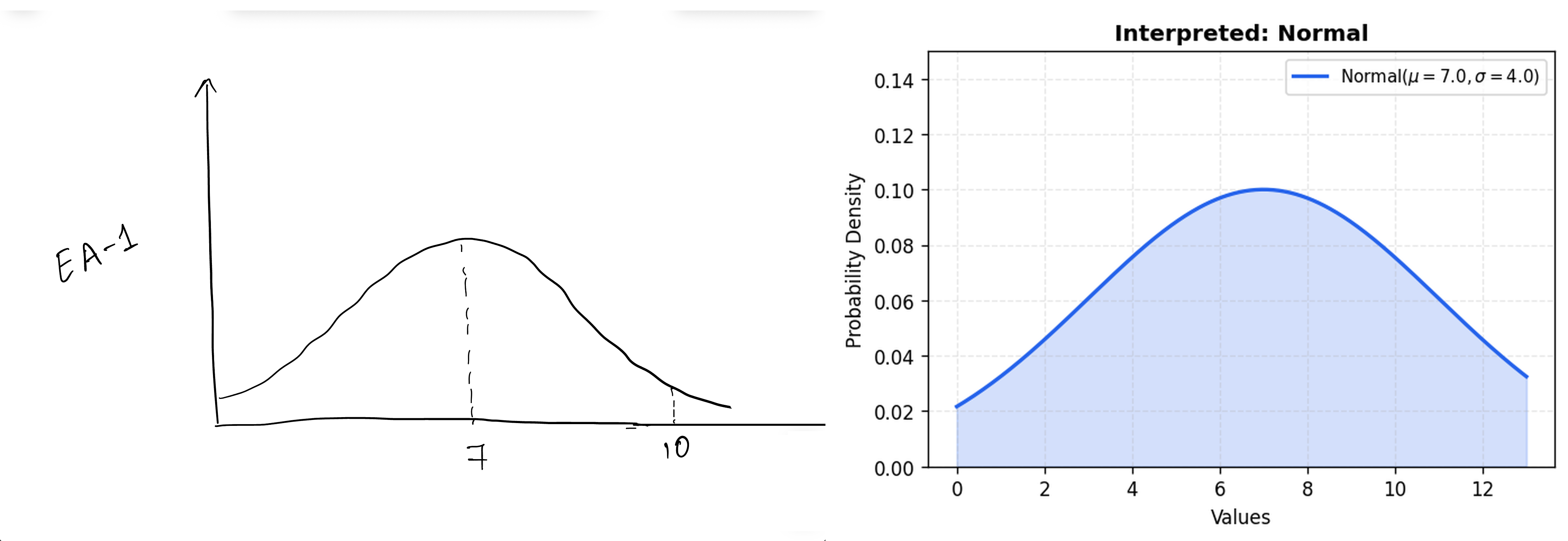}
    \caption{}
    \end{subfigure}
    \begin{subfigure}{0.8\textwidth}
\includegraphics[width=\linewidth]{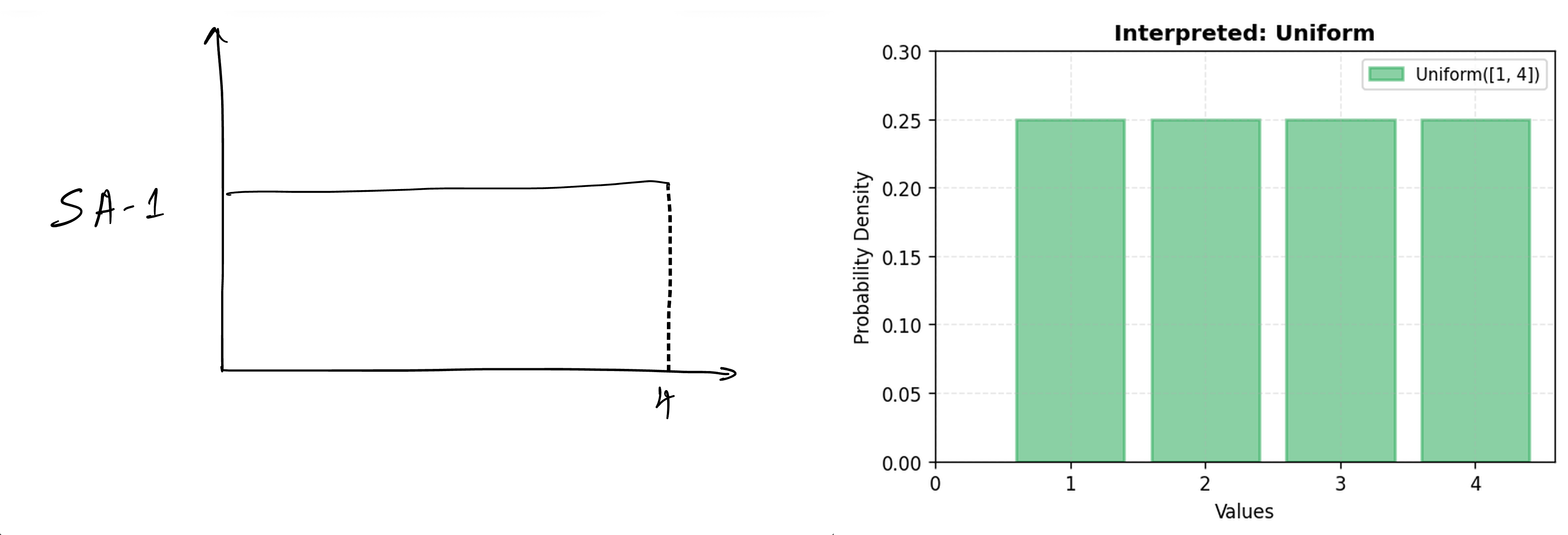}
        \caption{}
    \end{subfigure}
    \begin{subfigure}{0.8\textwidth}    
    \includegraphics[width=\linewidth]{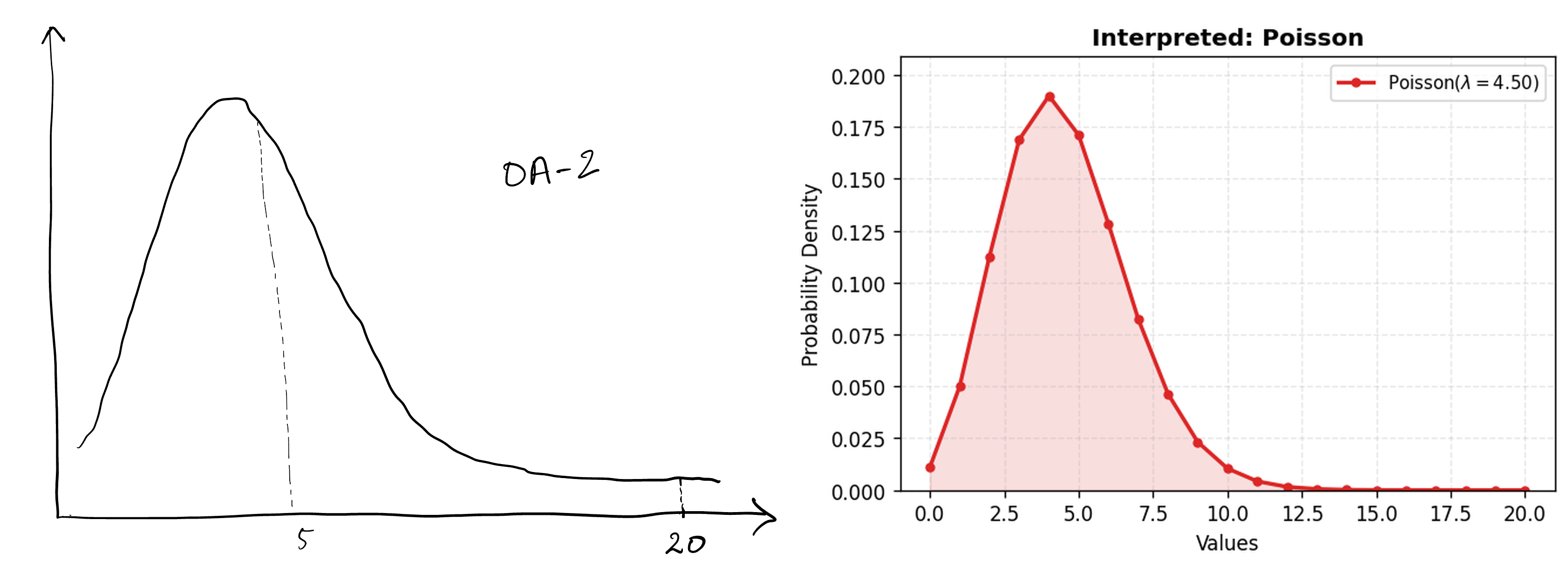}
    \caption{}
    \end{subfigure}
    \caption{Example hand-drawn images and corresponding LLM-generated distributions}
\label{fig:extracteddistr}
\end{figure*}
Users specify the parameters of the desired ABAC dataset through a JSON configuration file (input.json) with the following structure. Let \(|S|\) denote the number of subjects, \(|O|\) the number of objects, and \(|E|\) the number of environmental entities. The inputs are:

\begin{itemize}
\item Entity counts: \(n_{s}\) = \(|S|\), \(n_{o}\) = \(|O|\), \(n_{e}\) = \(|E|\) representing the number of subject, object and environment entities.

\item Attribute counts: \(n_{sa} = |A^s|\), \(n_{oa} = |A^o|\), \(n_{ea} =|A^e|\) for subject, object, and environmental attributes, respectively.
\item Attribute value cardinalities: \(n_{sa,i}, n_{oa,j}, n_{ea,k}\) representing the number of distinct values for each attribute, where for example, $n_{sa,1}$ denotes the number of possible values for the subject attribute $a_1^s$.
\item Probability distribution sets for attribute-value assignments: \(\mathcal{D}_{sa}, \mathcal{D}_{oa}, \mathcal{D}_{ea}\) for subject attributes, object attributes and environmental attributes, respectively. Each of these distributions capture how many of the corresponding entities can take each possible value of that attribute. 
\item Policy specification: number of permit rules $|\Pi^+|$ and deny rules $|\Pi^-|$.
\end{itemize}

The distributions currently supported by MuSimA include Sampled Normal (N) with parameters (\(\mu, \sigma^2\)), Poisson (P) with parameter \(\lambda\), and Uniform (U) distributions. These can be extended seamlessly without affecting the basic design of the tool. An example input JSON file is shown as Listing \ref{lst:inputjsonsample}. 



Considering the fact that the users may not always be fully familiar with the various distributions and their parameters, MuSimA supports a graphical input option as well. Instead of mentioning the distribution details in the input JSON as shown in Listing \ref{lst:inputjsonsample}, users can draw a sketch of the desired distribution of attribute values over entities. MuSimA uses an LLM to extract the distribution parameters that best fit the sketch. To handle potential inaccuracies in the initial extraction, MuSimA employs an iterative refinement process. Listing \ref{lst:refinementprompt} illustrates the prompt used to guide the LLM in refining the parameters by comparing the original sketch with a generated approximation.

After the distribution parameters are determined, MuSimA proceeds with the data generation process similar to text-based input. A few example input sketches and the corresponding extracted distributions are shown in Figure \ref{fig:extracteddistr}. The extraction logic is shown in Listing \ref{lst:vlmprompt}, and the end-to-end processing workflow is summarized in Sub-section \ref{subsec:handdrawn_pipeline}. Input images are sanitized using MIME type checks to prevent prompt injection. Note that, no LLM is involved in the text-only input mode.

\begin{lstlisting}[style=jsonbox,language=Python, caption={Iterative Refinement Prompt}, label={lst:refinementprompt}, basicstyle=\ttfamily\scriptsize, showstringspaces=false, breaklines=true]
        prompt_refine = f"""
        Refine the parameters. Left is Original, Right is Approximation.
        Current Parameters: {current_params}
        
        CRITICAL INSTRUCTION: 
        Look at the edges of the curve (at x_axis_min and x_axis_max).
        - If the Original sketch is still high at the edges, but your Approximation has dropped to zero, you MUST INCREASE SIGMA.
        - Do not sacrifice the width of the base just to make the peak sharper. 
        - It is better to have a wider curve that covers the edges than a narrow one that fits the peak perfectly.
        
        Output the CORRECTED object.
        """
\end{lstlisting}

\subsection{Visual Parameter Extraction}
\label{subsec:visual_param_extr}
For the multi-modal input, we utilize an LLM to extract distribution parameters from hand-drawn sketches. Listing \ref{lst:vlmprompt} shows the Python implementation of the extraction logic using the Gemini API.
\vspace{-0.1cm}
\begin{lstlisting}[style=jsonbox, language=Python, caption={LLM-based Parameter Extraction Logic}, label={lst:vlmprompt}, basicstyle=\ttfamily\scriptsize, showstringspaces=false, breaklines=true]
def analyze_and_refine(image_path):
    prompt_extract = """
    Analyze this image. Determine if it is a Normal, Poisson, or Uniform distribution.
    - For Normal: Identify peak (mu), width (sigma).
    - For Poisson: Identify rate (lambda).
    - For Uniform: Identify range (low, high).
    """
    
    # Call Gemini API
    response = client.models.generate_content(
        model="gemini-2.5-flash",
        contents=[prompt_extract, Image.open(image_path)],
        config={"response_mime_type": "application/json"}
    )
    
    params = parse_response(response)
    return params
\end{lstlisting}

\subsection{Hand-Drawn Input Processing Pipeline}
\label{subsec:handdrawn_pipeline}
The hand-drawn input pipeline processes a zip file (or folder) of sketches and produces a full ABAC configuration. For each image, it first uses the LLM to read the attribute identifier (SA/OA/EA and index) and the number of values from the sketch itself, avoiding reliance on filenames. It then classifies the distribution family (Normal, Poisson, or Uniform) and extracts parameters using a type-specific schema.
An interpreted plot is rendered from these parameters and compared against the original sketch. A refinement prompt iteratively corrects the parameters when necessary. The pipeline saves side-by-side comparison images and aggregates the extracted distributions and value counts into a JSON configuration. Finally, it enforces contiguous indices per attribute type (e.g., SA-1..SA-N) to ensure the output arrays align with attribute ordering.

\subsection{Dataset Generation Process}
\label{subsec:datasetgenproc}

The dataset generation process consists of the following stages:

\textbf{Stage 1: Entity and Attribute Generation.} Given the set of input specifications mentioned in Sub-section \ref{subsec:inputspec}, MuSimA generates the corresponding entity identifiers and attribute names as follows.
\begin{itemize}
\item \(S = \{s_1, s_2, \ldots, s_{n_{s}}\}\) as the set of subjects
\item \(O = \{o_1, o_2, \ldots, o_{n_{o}}\}\) as the set of objects
\item \(E = \{e_1, e_2, \ldots, e_{n_{e}}\}\) as the set of environmental entities
\item \(A^s = \{sa_1, sa_2, \ldots, sa_{n_{sa}}\}\) as the set of subject attributes
\item \(A^o = \{oa_1, oa_2, \ldots, oa_{n_{oa}}\}\) as the set of object attributes
\item \(A^e = \{ea_1, ea_2, \ldots, ea_{n_{ea}}\}\) as the set of environmental attributes
\end{itemize}

\textbf{Stage 2: Attribute Value Generation.} For each attribute \(a\), MuSimA generates a set of values \(V_a = \{a_1, a_2, \ldots, a_{v_a}\}\). These are stored in mapping structures \textit{SAV}, \textit{OAV} and \textit{EAV}, respectively, where \(SAV[a_i^s]\) contains all possible values for the subject attribute \(a_i^s\).

\textbf{Stage 3: Entity-Attribute-Value Assignment.} For each entity \(e \in S \cup O \cup E\), MuSimA assigns attribute values for each relevant attribute according to the specified distributions. Let \(a\) be an attribute with \(n\) possible values \((a_1, a_2, \ldots, a_n)\), and \(\mathcal{D}\) denote the desired distribution for that attribute. Depending on the nature of the distribution, attribute-values are assigned as described below.


\begin{itemize}
    \item \textbf{Normal Distribution (N):} MuSimA samples a real number \(x\) from a truncated normal distribution on \([0, n]\) with user-specified mean ($\mu$) and variance ($\sigma^2$), and then assigns value \(a_k\) if
    \[
    x \in [k-1, k), \quad \text{for } k = 1, 2, \ldots, n.
    \]
    This bins the interval \([0, n]\) into \(n\) unit ranges and maps the sampled value to an attribute value accordingly.
    
    \item \textbf{Poisson Distribution (P):} Given parameter \(\lambda\), MuSimA computes probabilities 
    \[
    P(X = k) = \frac{\lambda^k e^{-\lambda}}{k!}, \quad \text{for } k = 1, 2, \ldots, n,
    \]
    normalizes them so that the weights sum to 1, and samples \(a_k\) according to these discrete probabilities.
    
    \item \textbf{Uniform Distribution (U):} Since all attribute values are equally likely, the assignment samples an index \(k \in [1, n]\) uniformly at random and assigns \(a_k\).
\end{itemize}

Thus, for each attribute assignment:
\[
v \sim \mathcal{D}, \qquad v \in V_a = \{a_1, a_2, \ldots, a_n\}
\]

These assignments are stored in mapping structures: \(SV[s_i] = [v_1, v_2, \ldots, v_{n_{sa}}]\) for subjects, \(OV[o_j] = [v_1, v_2, \ldots, v_{n_{oa}}]\) for objects, and \(EV[e_k] = [v_1, v_2, \ldots, v_{n_{ea}}]\) for environmental entities.

\textbf{Stage 4: ABAC Policy Generation.} MuSimA generates $|\Pi^+|$ permit rules and $|\Pi^-|$ deny rules. Each permit rule $r_i^+ \in \Pi^+$ and deny rule $r_i^- \in \Pi^-$ is of the form:

\[
r_i^{\pm}: \bigwedge_{j=1}^{n_{sa}} (sa_j = v_j^{sa}) \land \bigwedge_{k=1}^{n_{oa}} (oa_k = v_k^{oa}) \land \bigwedge_{l=1}^{n_{ea}} (ea_l = v_l^{ea}) \rightarrow \text{permit} / \text{deny}
\]

where each condition specifies an attribute-value pair. Multiple permit rules combine disjunctively (access granted if \textit{any} permit rule matches). However, in case of conflicting policies,  deny rules take precedence (deny-overrides-permit semantics).

\textbf{Stage 5: Access Control Matrix Generation.} For each tuple \((s_i, o_j, e_k) \in S \times O \times E\), MuSimA evaluates the attribute values against both rule sets. An entry in the access control matrix \(ACM[s_i, o_j, e_k]\) is set to 1 (permit) if any permit rule in $\Pi^+$ matches and no deny rule in $\Pi^-$ matches, and 0 (deny) otherwise, i.e.,

\[
ACM[s_i, o_j, e_k] = \begin{cases}
1 & \text{if } \exists r_i^+ \in \Pi^+: \text{conditions satisfied} \\
  & \quad \land \nexists r_j^- \in \Pi^-: \text{conditions satisfied} \\
0 & \text{otherwise}
\end{cases}
\]

\begin{lstlisting}[style=jsonbox, caption={Output JSON for Input of Listing \ref{lst:inputjsonsample}}, label={lst:outputjsonsample}]
{
  "S": ["S_1", "S_2", "S_3"], "O": ["O_1", "O_2", "O_3"], "E": ["E_1", "E_2"],
  "SA": ["SA_1", "SA_2"], "OA": ["OA_1", "OA_2"], "EA": ["EA_1"],
  "SAV": {
    "SA_1": ["SA_1_1", "SA_1_2"], 
    "SA_2": ["SA_2_1", "SA_2_2", "SA_2_3", "SA_2_4"]
  },
  "OAV": { "OA_1": ["OA_1_1", "OA_1_2"], "OA_2": ["OA_2_1"] },
  "EAV": { "EA_1": ["EA_1_1", "EA_1_2"] },
  "SV": {
    "S_1": ["SA_1_2", "SA_2_3"], "S_2": ["SA_1_2", "SA_2_4"], 
    "S_3": ["SA_1_1", "SA_2_3"]
  },
  "OV": {
    "O_1": ["OA_1_2", "OA_2_1"], "O_2": ["OA_1_1", "OA_2_1"],
    "O_3": ["OA_1_1", "OA_2_1"]
  },
  "EV": { "E_1": ["EA_1_2"], "E_2": ["EA_1_2"] },
  "permit_rules": [
    "SA_1=SA_1_2, SA_2=SA_2_4, OA_1=OA_1_1, OA_2=OA_2_1, EA_1=EA_1_2"
  ],
  "deny_rules": [
    "SA_1=SA_1_1, SA_2=SA_2_3, OA_1=OA_1_1, OA_2=OA_2_1, EA_1=EA_1_2"
  ],
}
\end{lstlisting}

\subsection{Output File Formats}
\label{subsec:outputfileformat}

MuSimA generates output files representing different components of the generated ABAC dataset:

\textbf{i. output.json:} Contains the structured representation of the complete ABAC system, including (Listing~\ref{lst:outputjsonsample}):
\begin{itemize}
\item Entity sets: \(S, O, E\) with identifiers
\item Attribute definitions: \(A^s, A^o, A^e\)
\item Attribute value sets: \(SAV, OAV, EAV\) mapping each attribute to its possible values
\item Entity-attribute assignments: \(SV, OV, EV\) mapping each entity to its assigned attribute values
\item A permit policy $\Pi^+$ and a deny policy $\Pi^-$: The number of rules in each is the same as that specified by \texttt{accepted\_rules\\\_count} and \texttt{denied\_rules\_count} in input.json.
\end{itemize}

\textbf{ii. ACM.txt:} Contains the Access Control Matrix with decision entries (0/1) for all $(s_i, o_j, e_k)$ combinations.





\textbf{iii. access\_data.txt:} A flattened representation where each line contains the attribute values for a subject-object-environment tuple and the corresponding access decision $d \in \{0, 1\}$, suitable for machine learning applications.


\textbf{iv. Distribution Attestation:} A set of images (similar to Figure \ref{fig:Detaileddistribution} shown later in Section \ref{sec:exprslt}) depicting how well the generated output distributions match the desired distributions. For multi-modal input, additional images are produced plotting the extracted distributions alongside the hand-drawn distributions (similar to Figure \ref{fig:extracteddistr}).

\section{Experimental Results}
\label{sec:exprslt}






To validate the correctness of the attribute value assignment process, we computed the expected count for each attribute value under different distributions and compared them against actual generated counts.
For each attribute $i$ with $n_i$ possible values $\{a_1, a_2, \ldots, a_{n_i}\}$, we define the attribute-level error as:
\[
\epsilon_i = \sum_{k=1}^{n_i} \frac{|E[\text{Count}(a_k)] - \text{Actual}(a_k)|}{\text{Actual}(a_k)}
\]
Where, $E[\text{Count}(a_k)]$, the expected count for each $a_k$ is computed from the corresponding distribution, i.e., Uniform, truncated Normal, or normalized Poisson (See Section \ref{subsec:datasetgenproc}). 
For each distribution type, the average error across all attributes is computed as:
\[
\bar{\epsilon} = \frac{1}{N} \sum_{i=1}^{N} \epsilon_i
\]
where \(N\) is the number of attributes for that distribution.

\begin{figure*}
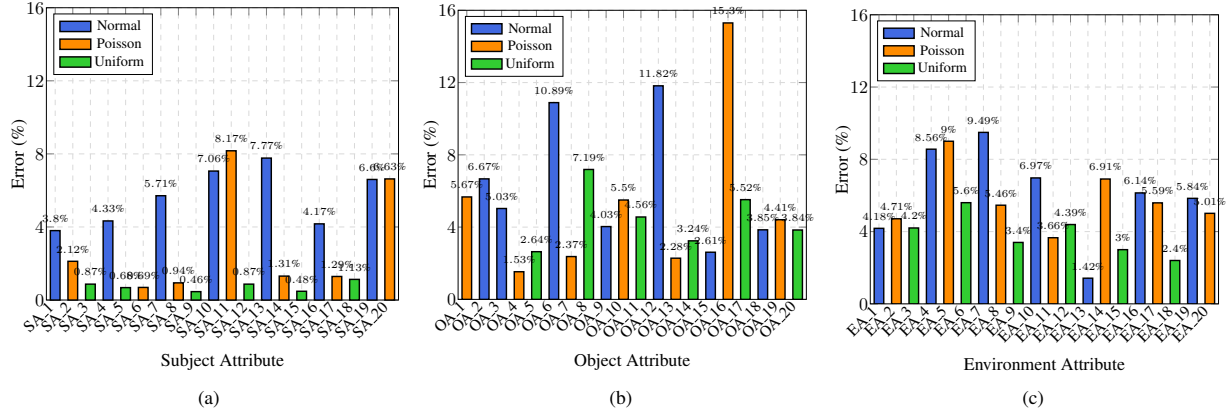

\advance\leftskip-0.00cm
 \resizebox{1\textwidth}{!}{
\centering
\begin{subfigure}[b]{.4\textwidth}
    \resizebox{\textwidth}{!}{%
\errorDistributionChart{SA}{Subject Attribute}{}{3.25}{16}{0,4,8,12,16}
{
 (1,3.80) (4,4.33) (7,5.71) (10,7.06) 
 (13,7.77) (16,4.17) (19,6.60)
}
{
 (2,2.12) (6,0.69) (8,0.94) 
 (11,8.17) 
 (14,1.31) (17,1.29) (20,6.63)
}
{
 (3,0.87) (5,0.68) (9,0.46) (12,0.87) 
 (15,0.48) (18,1.13)
}}
\caption{}
\label{fig:sa_errors}
 \end{subfigure}
 \hfill
 \begin{subfigure}[b]{0.4\textwidth}
    \resizebox{\textwidth}{!}{%
 \errorDistributionChart{OA}{Object Attribute}{}{5.45}{16}{0,4,8,12,16}
{
 (2,6.67) (3,5.03) (6,10.89) (9,4.03) (12,11.82) (15,2.61) (18,3.85)
}
{
 (1,5.67) (4,1.53) (7,2.37) (10,5.50) (13,2.28) (16,15.30) (19,4.41)
}
{
 (5,2.64) (8,7.19)  (11,4.56) (14,3.24) (17,5.52) (20,3.84)
}
}
\caption{}
\label{fig:oa_errors}
 \end{subfigure}
 \hfill
 \begin{subfigure}[b]{0.4\textwidth}
    \resizebox{\textwidth}{!}{%
\errorDistributionChart{EA}{Environment Attribute}{}{5.30}{16}{0,4,8,12,16}
{
 (1,4.18) (4,8.56) (7,9.49) (10,6.97) (13,1.42) (16,6.14) (19,5.84)
}
{
 (2,4.71) (5,9.00) (8,5.46) (11,3.66) (14,6.91) (17,5.59) (20,5.01)
}
{
 (3,4.20) (6,5.60)  (9,3.40) (12,4.39) (15,3.00) (18,2.40)
}
}
\caption{}
\label{fig:ea_errors}
 \end{subfigure}
 \hfill
 
 }
 \caption{Error Distribution 
 for (a) Subject Attribute (SA) (b) Object Attribute (OA) (c) Environment Attribute (EA).}
 \label{fig:ExperimentalResults}
\end{figure*}

\begin{figure*}
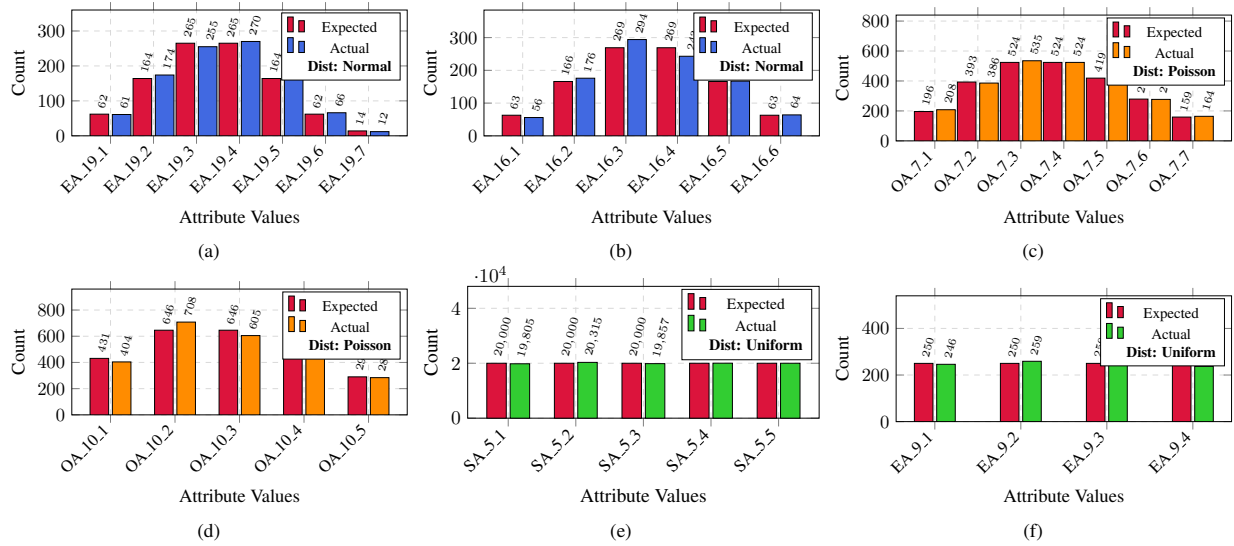

\advance\leftskip-0.00cm

\resizebox{1\textwidth}{!}{
\centering
\begin{subfigure}[b]{0.4\textwidth}
\resizebox{\textwidth}{!}{%
\countComparisonChart{EA\_19}{Normal}{distNormal}{300}
{
 (1,62) (2,164) (3,265) (4,265) (5,164) (6,62) (7,14)
}
{
 (1,61) (2,174) (3,255) (4,270) (5,162) (6,66) (7,12)
}
{
EA\_19\_1, EA\_19\_2, EA\_19\_3, EA\_19\_4, EA\_19\_5, EA\_19\_6, EA\_19\_7}
}
\caption{}
\label{fig:ea19_counts}
\end{subfigure}


\begin{subfigure}[b]{0.4\textwidth}
\resizebox{\textwidth}{!}{%
\countComparisonChart{EA\_16}{Normal}{distNormal}{320}
{
 (1,63) (2,166) (3,269) (4,269) (5,166) (6,63)
}
{
 (1,56) (2,176) (3,294) (4,243) (5,167) (6,64)
}
{EA\_16\_1, EA\_16\_2, EA\_16\_3, EA\_16\_4, EA\_16\_5, EA\_16\_6}
}
\caption{}
\label{fig:ea16_counts}
\end{subfigure}

\begin{subfigure}[b]{0.4\textwidth}
\resizebox{\textwidth}{!}{%
\countComparisonChart{OA\_7}{Poisson}{distPoisson}{700}
{
 (1,196) (2,393) (3,524) (4,524) (5,419) (6,279) (7,159)
}
{
 (1,208) (2,386) (3,535) (4,524) (5,406) (6,277) (7,164)
}
{OA\_7\_1, OA\_7\_2, OA\_7\_3, OA\_7\_4, OA\_7\_5, OA\_7\_6, OA\_7\_7}
}
\caption{}
\label{fig:oa7_counts}
\end{subfigure}
 }

\resizebox{1\textwidth}{!}{
\centering
\begin{subfigure}[b]{0.4\textwidth}
\resizebox{\textwidth}{!}{%
\countComparisonChart{OA\_10}{Poisson}{distPoisson}{800}
{
 (1,431) (2,646) (3,646) (4,484) (5,290)
}
{
 (1,404) (2,708) (3,605) (4,499) (5,284)
}
{OA\_10\_1, OA\_10\_2, OA\_10\_3, OA\_10\_4, OA\_10\_5}
}
\caption{}
\label{fig:oa10_counts}
\end{subfigure}

\begin{subfigure}[b]{0.4\textwidth}
    \resizebox{\textwidth}{!}{%
\countComparisonChart{SA\_5}{Uniform}{distUniform}{40000}
{
 (1,20000) (2,20000) (3,20000) (4,20000) (5,20000)
}
{
 (1,19805) (2,20315) (3,19857) (4,20023) (5,20000)
}
{SA\_5\_1, SA\_5\_2, SA\_5\_3, SA\_5\_4, SA\_5\_5}
}
\caption{}
\label{fig:sa5_counts}
  \end{subfigure}

\begin{subfigure}[b]{0.4\textwidth}
\resizebox{\textwidth}{!}{%
\countComparisonChart{EA\_9}{Uniform}{distUniform}{450}
{
 (1,250) (2,250) (3,250) (4,250) 
}
{
 (1,246) (2,259) (3,258) (4,237) 
}
{EA\_9\_1, EA\_9\_2, EA\_9\_3, EA\_9\_4}
}
\caption{}
\label{fig:ea9_counts}
\end{subfigure}
 }

\caption{Expected vs Actual Count for (a) EA\_19 (b) EA\_16 (c) OA\_7 (d) OA\_10 (e) SA\_5 (f) EA\_9}
\label{fig:Detaileddistribution}
\end{figure*}

To evaluate the performance of the MuSimA in generating large-scale datasets, we conducted experiments with the following configuration: 100,000 subjects, 2500 objects, 1000 environmental entities, and 50 authorization rules. A total of 60 distributions were used (20 for subject attributes, 20 for object attributes, and 20 for environmental attributes). Table~\ref{tab:large_scale_errors} summarizes the attribute-level mean errors for each entity type while Figures \ref{fig:ExperimentalResults}(a)-(c)
show the detailed per-attribute error distribution for subject, object, and environment attributes respectively. Subject attributes exhibit the lowest errors on average (3.25\%), with most attributes showing errors below 4\%. Object attributes display  errors averaging 5.45\%, while environment attributes show  errors at 5.30\%. The higher error in OA\_16 (15.3\%) indicates a particularly challenging distribution.

\begin{table}[t]
\centering
\caption{Mean Error by Entity Type for Large-Scale Dataset}
\label{tab:large_scale_errors}
\resizebox{0.7\columnwidth}{!}{
\begin{tabular}{lcc}
\hline
\textbf{Entity Type} & \textbf{No. of Attributes} & \textbf{Mean Error (\%)} \\
\hline
Subject Attributes (SA) & 20 & 3.25 \\
Object Attributes (OA) & 20 & 5.45 \\
Environment Attributes (EA) & 20 & 5.30 \\
\hline
\textbf{Overall} & \textbf{60} & \textbf{4.66} \\
\hline
\end{tabular}
}
\end{table}

To validate the correctness of the data generation process, we analyzed the specific value counts for selected attributes. The comparisons between Actual vs. Expected counts for the Normal distribution (EA\_19 and EA\_16) are shown in Figures \ref{fig:ea19_counts} and \ref{fig:ea16_counts}. For the Poisson distribution (OA\_7 and OA\_10), the results are presented in Figures \ref{fig:oa7_counts} and \ref{fig:oa10_counts}. Finally, the results for the Uniform distribution (SA\_5 and EA\_9) are illustrated in Figures \ref{fig:sa5_counts} and \ref{fig:ea9_counts}.
Attributes with simpler distributions (e.g., uniform or near-uniform) tend to have lower errors, while those with complex or highly skewed distributions show higher errors. Despite this variation, the overall mean error of 4.66\% confirms that the tool maintains acceptable accuracy even when generating large-scale datasets with thousands of entities.

\section{Related Work}
\label{sec:related}
Over the last few years, ABAC has emerged as one of the most researched access control models for enforcing organizational security policies with efforts to improve policy mining algorithms~\cite{xustoller}, extraction of machine-enforceable policies from natural language specifications~\cite{MianSACMAT2025}, and enhance scalability~\cite{cosequeue}. ABAC has found significant applications in healthcare,  university systems, cloud computing~\cite{Weng2025}, IoT~\cite{abaciot1} and blockchain-based systems~\cite{DBLP:journals/cn/DingWMD25}.

ABAC policy mining aims to automate the transition to ABAC by generating policies from existing lower-level access control data~\cite{XuStoller2018}. Xu and Stoller~\cite{xustoller} introduced the first ABAC policy mining algorithm whose datasets---including University, Project Management, and Healthcare scenarios---have become common benchmarks in the field. Subsequent works have developed evolutionary algorithms for mining both permit and deny rules. However, existing datasets highlight limitations, particularly regarding the proportion of access requests versus sparse logs, necessitating synthetic log generation capabilities~\cite{Mocanu2015}.

Two publicly available real-world datasets are the Amazon Kaggle~\cite{Amazon_dataset} and Amazon UCI~\cite{AmazonUciDataset} datasets, commonly used by researchers. The Amazon Kaggle dataset provides only eight anonymized user attributes with no resource attributes and lacks a rule set essential for comprehensive evaluation. The Amazon UCI dataset similarly lacks clear attribute data and rule sets, with cases where identical users have different access permissions. Both datasets are unsuitable for generating interpretable ABAC rules with reasonable accuracy~\cite{Rai2023}.

Existing synthetic generation approaches have significant limitations. ConGRASS~\cite{Rai2023} uses Conditional Tabular GAN to generate data but remains constrained by the original datasets' attribute dimensionality. Xu and Stoller's simulated datasets~\cite{xustoller} allow only minor pre-determined variations, cannot generate domain-specific distributions, and cannot handle environmental attributes---a major ABAC differentiator from RBAC and DAC.

ABAC Lab~\cite{ABACLab} is an interactive platform providing a repository of datasets including the E-Document dataset and Workforce Management dataset. It offers synthetic log generation, rule analysis, and visualization features. However, it focuses primarily on policy mining rather than flexible dataset generation, using pseudo-random algorithms with pre-configured size parameters instead of allowing users to specify arbitrary attribute distributions.


Table~\ref{tab:comparison} provides a qualitative comparison of MuSimA with existing tools and datasets highlighting the key differentiating features.

\begin{table}[t]
\centering
\caption{Qualitative Comparison of ABAC Data Generation Tools}
\label{tab:comparison}
\resizebox{\columnwidth}{!}{
\begin{tabular}{lcccc}
\hline
\textbf{Feature} & \textbf{Xu \& Stoller \cite{xustoller}} & \textbf{ConGRASS \cite{Rai2023}}       & \textbf{ABAC Lab \cite{ABACLab}}       & \textbf{MuSimA \textbf{This work}}         \\
\hline
Env. attributes      & \texttimes & \texttimes & \texttimes & \checkmark \\
User-specified dist. & \texttimes & \texttimes & \texttimes & \checkmark \\
Multi-modal input    & \texttimes & \texttimes & \texttimes & \checkmark \\
Arbitrary scale      & \texttimes & \checkmark & \texttimes & \checkmark \\
Deny rules           & \checkmark & \texttimes & \texttimes & \checkmark \\
ACM generation       & \texttimes & \texttimes & \texttimes & \checkmark \\
Open source          & \checkmark & \checkmark & \checkmark & \checkmark \\
\hline
\end{tabular}%
}
\end{table}

\section{Conclusion and Future Directions}
\label{sec:concl}

We have proposed MuSimA - an ABAC data generator with multi-modal input (text-based JSON or hand-drawn distribution sketches). The output contains the details of the intended ABAC dataset along with distribution attestation images. While MuSimA is a web based tool, 
due to its modular design, it supports \textit{Bring Your Own Model}, and the source code is open source\footnote{\url{https://doi.org/10.5281/zenodo.18876163}} for organizations to run the tool locally for avoiding potential privacy concerns.

There are several promising directions for future development.
Instead of manually specifying all parameters, future versions could analyze real access logs or organizational data to learn realistic patterns, extract attribute distributions, and identify correlations between attributes. Modeling such correlated attribute distributions which is not supported now, would produce datasets that better reflect real-world scenarios and is an important future extension of this work.
MuSimA currently supports both permit and deny rules with deny-overrides semantics. Advanced policy features such as conditional policies based on time or location, role hierarchies, and separation-of-duty constraints are planned.
Comparison of the generated datasets with existing real-world datasets like Amazon Kaggle~\cite{Amazon_dataset} and Amazon UCI~\cite{AmazonUciDataset} to evaluate the realistic nature of the datasets generated by MuSimA is another promising direction for future work.

\printbibliography
\end{document}